\documentclass{article} 
\usepackage{iclr2027_conference,times}

\usepackage{amsmath,amsfonts,bm}

\def\eqref#1{equation~\ref{#1}}

\def\1{\bm{1}}

\DeclareMathAlphabet{\mathsfit}{\encodingdefault}{\sfdefault}{m}{sl}
\SetMathAlphabet{\mathsfit}{bold}{\encodingdefault}{\sfdefault}{bx}{n}

\usepackage{url}
\usepackage{graphicx}
\usepackage{booktabs}
\usepackage{pifont}
\usepackage{amsfonts}
\usepackage{amssymb}
\usepackage{bbm}
\usepackage[table]{xcolor}
\usepackage{wrapfig}
\usepackage{caption}
\usepackage{subcaption}
\usepackage[most]{tcolorbox}
\usepackage{multicol}
\usepackage{multirow}
\usepackage{array}
\usepackage{amsmath}
\usepackage{fontawesome5}
\usepackage[hidelinks]{hyperref}

\definecolor{rowgray}{HTML}{F7F7F7}
\definecolor{checkgreen}{HTML}{2A9D6F}
\definecolor{crossred}{HTML}{D05A5A}
\newcommand{\goodmark}{\textcolor{checkgreen}{\ding{51}}}
\newcommand{\badmark}{\textcolor{crossred}{\ding{55}}}
\definecolor{questionborder}{HTML}{A8B6C8}
\definecolor{questionbg}{HTML}{F6F8FB}
\definecolor{questiontext}{HTML}{334155}
\definecolor{lowrt}{RGB}{222,242,222}
\definecolor{highrt}{RGB}{255,230,230}
\definecolor{visgreen}{RGB}{239,248,241}
\definecolor{visgreenstrong}{RGB}{220,241,225}
\definecolor{stdgray}{RGB}{105,105,105}
\definecolor{darkred}{RGB}{120,45,45}
\newcolumntype{G}{>{\columncolor{visgreen}}c}
\newcommand{\attackcell}[2]{#1{\fontsize{6}{6}\selectfont\color{stdgray}\,$\pm$#2}}
\newcommand{\viscell}[2]{\textbf{#1}{\fontsize{6}{6}\selectfont\color{stdgray}\,$\pm$#2}}

\newcommand{\githuburl}{https://github.com/holdrain/Coverlock}
\newcommand{\projecturl}{https://holdrain.github.io/Coverlock}

\title{Residual Transferability in Neural Image\\Watermarking}

\makeatletter
\renewcommand{\@maketitle}{%
  \vbox{%
    \hsize\textwidth
    \centering
    {\LARGE\scshape \@title\par}%
    \vspace{0.16in}%
    {\normalsize
      \begin{tabular}{@{}ccc@{}}
        \textbf{Ziping Dong}$^{1}$ &
        \hspace{0.65cm}\textbf{Qi Li}$^{1}$\hspace{0.65cm} &
        \textbf{Xinchao Wang}$^{1,*}$
      \end{tabular}\par
    }%
    \vspace{5pt}%
    {\small $^{1}$National University of Singapore\par}%
    \vspace{3pt}%
    {\footnotesize
      \texttt{zipingdong@.u.nus.edu}\quad
      \texttt{liqi@u.nus.edu}\quad
      \texttt{xinchao@nus.edu.sg}\par
    }%
    \vspace{2pt}%
    {\footnotesize $^{*}$Corresponding author\par}%
    \vspace{4pt}%
    {\small
      \href{\githuburl}{\faGithub\,\textbf{GitHub}}%
      \hspace{1.1cm}%
      \href{\projecturl}{\faGlobe\,\textbf{Project Page}}\par
    }%
    \vspace{4pt}%
  }%
}
\makeatother

\iclrfinalcopy

\begin{document}
\maketitle

\begin{abstract}
Neural image watermarks can be forged by extracting watermark-bearing residuals from released images and transferring them to unrelated content. While prior work has demonstrated this vulnerability, what makes these residuals transferable remains poorly understood. We formalize this vulnerability with \textbf{residual transferability (RT)}, a metric that quantifies how well watermark evidence remains decodable after transfer across unrelated images. Through comparative analyses and controlled interventions, we find that common training-side variations do not account for the large RT differences across watermarking systems; instead, architectural design plays a central role. By contrasting high- and low-RT systems and validating their architectural differences through controlled interventions, we identify two mechanisms that strengthen the dependence of watermark evidence on the cover image, thereby suppressing the residual transferability.
These findings provide concrete design guidance for developing more forgery-resistant watermarking architectures. Complementarily, for existing watermarking systems where architectural redesign is impractical, we introduce \textbf{CoverLock}, a plug-and-play strategy for existing watermarking systems that strengthens such image dependence without architectural redesign. Across representative watermarking systems exhibiting high residual transferability, CoverLock achieves a more favorable security--robustness trade-off than both traditional handcrafted defenses and learned classifier-based defenses.
\end{abstract}

\section{Introduction}

The growing prevalence of AI-generated images has made invisible watermarking~\citep{zhu2018hidden, RivaGAN,stablesignature, VINE, soucek2025pixelseal} increasingly important for content provenance and attribution.
By embedding imperceptible message-carrying perturbations, watermarking systems enable ownership verification and source attribution through subsequent decoding. Major technology companies, including Google (SynthID)~\citep{Synthid}, Meta~\citep{metawatermark}, OpenAI~\citep{openaiwatermark}, and Amazon~\citep{Amazonwatermark}, have already integrated watermarking solutions into their generative AI ecosystems to improve transparency, trace harmful or misleading AI-generated content, and support content accountability.

However, recent studies have shown that current image watermarking systems remain vulnerable to security threats.
Among these threats, residual-based forgery attack~\citep{muller2024black,dong2025wmcopier,souvcek2026transferable} has drawn increasing attention, as attackers can transfer a watermark from legitimate images to unauthorized content and induce false attribution. Recent studies further show that such attacks may require only one or a few watermarked images~\citep{yang2024steganalysis,souvcek2026transferable}, substantially reducing the barrier to forgery and challenging the reliability of watermark-based attribution.
Such attacks recover watermark-bearing residuals through different estimation procedures, averaging a collection of watermarked images with natural-image references~\citep{yang2024steganalysis} or using an auxiliary reference model~\citep{souvcek2026transferable}.

Despite their demonstrated effectiveness and low attack cost, these studies primarily focus on constructing increasingly effective forgery attacks, while the intrinsic property that enables such cross-image transfer remains poorly understood. Elucidating the mechanisms underpinning the success of such attacks is essential for both developing principled defenses and informing the design of more secure watermarking systems.

To this end, we first formalize residual transferability (RT) as an inherent property of watermarking systems: watermark residuals may remain decodable after being transferred across unrelated cover images. We then quantify this property using the RT metric, which measures a system's susceptibility to residual-based forgery independently of any specific residual estimation procedure. 
As summarized in Table~\ref{tab:watermarking_comparison}, we select five representative watermarking schemes spanning different training configurations and architectural designs, and observe strikingly different levels of RT. Even methods trained with similar objectives and robustness-oriented augmentations can exhibit substantially different transferability. These observations motivate the following question:

\begin{table*}[t]
\centering
\caption{
Comparison of representative watermarking methods in terms of training configurations,
architectural designs, and residual transferability.
Lower RT indicates a lower risk of residual-based watermark forgery.
The computation of RT@100 is detailed in Sec.~\ref{sec:rt_measurement}.
}
\label{tab:watermarking_comparison}

\small
\setlength{\tabcolsep}{2.5pt}
\renewcommand{\arraystretch}{1.06}

\resizebox{\textwidth}{!}{%
\begin{tabular}{l l c c c l c c c}
\toprule
Method & Loss & Disc. & Noise & Arch. & Dataset
& Resolution & Message Len. & RT \\
\midrule
CIN~\citep{CIN}
& MSE+MSE
& \badmark
& \goodmark
& INN
& COCO
& 128
& 30
& 1.00 \\

VINE~\citep{VINE}
& BCE+MSE+LPIPS
& \goodmark
& \goodmark
& CNN
& OpenImages
& 512
& 100
& 0.99 \\

MBRS~\citep{jia2021mbrs}
& MSE+MSE
& \goodmark
& \goodmark
& CNN
& ImageNet
& 128
& 256
& 0.98 \\

\midrule
\rowcolor{rowgray}
HiDDeN~\citep{zhu2018hidden}
& MSE+MSE
& \goodmark
& \goodmark
& CNN
& COCO
& 128
& 30
& \textbf{0.28} \\

\rowcolor{rowgray}
RivaGAN~\citep{RivaGAN}
& BCE+Feature
& \goodmark
& \goodmark
& CNN
& Hollywood2
& 256
& 32
& \textbf{0.08} \\
\bottomrule
\end{tabular}%
}
\end{table*}


\begin{center}
\setlength{\fboxrule}{0.7pt}
\setlength{\fboxsep}{8pt}
\fcolorbox{gray!80!black}{gray!2}{%
\parbox{0.90\linewidth}{%
\centering
\color{black!95}
\bfseries
What makes watermark residuals transferable across unrelated images?
}}
\end{center}

To answer this question, we first conduct a comparative behavioral analysis across all evaluated watermarking systems, examining how watermark information is encoded and represented by their encoders and decoders. This helps us narrow down the plausible explanations for RT. 
RT reveals a clear distinction in the relationship between watermark information and image content: systems with high RT encode and decode watermark signals with limited reliance on the host image, whereas systems with low RT retain a strong coupling between watermark evidence and image content. This observation points to a possible shortcut mechanism: the model learns to encode watermark information through a largely \emph{cover-independent pathway}, which in turn makes the residual transferable across unrelated images.

This interpretation naturally narrows the search space for understanding why some systems exhibit low RT. Through controlled interventions on low-RT systems, we identify two architectural mechanisms that strengthen the dependence between watermark information and image content, thereby substantially reducing residual transferability. These findings reveal concrete architectural principles for suppressing cover-independent watermark pathways and improving resistance to residual-based forgery. 

Architectural redesign, however, can be method-specific and costly for existing watermarking systems. As a complementary option, we introduce \textbf{CoverLock}, a lightweight image-binding wrapper that incorporates image-dependent visual information into the decoding process without redesigning the core watermark architecture. Across three high-RT systems and both existing residual-based forgery attacks, CoverLock achieves a more favorable security--robustness trade-off than prior defenses. Our main contributions are summarized as follows.

\begin{itemize}
\item We introduce residual transferability (RT), a system-level property that measures the cross-image decodability of watermark residuals and characterizes susceptibility to residual-based forgery independent of the specific residual estimation method.

\item Through controlled studies of training and architectural factors, we identify model architecture as a major factor governing RT. We further isolate two architectural design choices that substantially reduce residual transferability.

\item As a complementary option for existing systems, we introduce \textbf{CoverLock}, an image-binding wrapper that mitigates RT without redesigning the watermark architecture. Across three high-RT systems, CoverLock achieves a more favorable security--robustness trade-off than prior defenses, reducing average forgery success to below 0.25\% under both existing residual-based forgery attacks while attaining higher robustness.

\end{itemize}

\section{Background and Problem Formulation}

\subsection{Image Watermarking}

Given an image $x\in\mathbb{R}^{H\times W\times 3}$ of height $H$ and width $W$, and a $L$-bit binary message $m\in\{0,1\}^{L}$, a watermark encoder $E$ embeds $m$ into $x$, while a decoder $D$ recovers the message from the resulting watermarked image:
\begin{equation}
x^{w}=E(x,m), \qquad m'=D\bigl(x^{w}\bigr).
\end{equation}
Here, $x^{w}$ denotes the watermarked image, $m'$ is the recovered message. 
During training, the encoder and decoder are typically optimized jointly using
\begin{equation}
\mathcal{L}
=
\lambda_{\mathrm{wm}}
\ell_{\mathrm{wm}}
\bigl(D(\tau(x^{w})),m\bigr)
+
\lambda_{\mathrm{img}}
\ell_{\mathrm{img}}(x^{w},x),
\end{equation}
where $\ell_{\mathrm{wm}}$ and $\ell_{\mathrm{img}}$ denote the message-recovery and image-fidelity losses, respectively, $\lambda_{\mathrm{wm}}$ and $\lambda_{\mathrm{img}}$ are their corresponding weights, and $\tau\sim\mathcal{T}$ denotes a sampled transformation used for robustness training~\citep{zhu2018hidden,satheesh2026compositionaladversarialtrainingrobust}.

\subsection{Residual-based Watermark Forgery}
\label{sec:residual_forgery}

Residual-based watermark forgery~\citep{yang2024steganalysis,souvcek2026transferable} aims to make an unrelated image appear to carry the watermark of released watermarked images. Given $s$ reference watermarked images $\{x_{r,i}^{w}\}_{i=1}^{s}$ carrying the same message $m$, the attacker recovers approximate clean counterparts $\{\hat{x}_{r,i}\}_{i=1}^{s}$ and aggregates the estimated watermark residuals:
\begin{equation}
\hat{\bar{r}}_{s}
=
\frac{1}{s}
\sum_{i=1}^{s}
\left(
x_{r,i}^{w}-\hat{x}_{r,i}
\right).
\label{eq:estimated_residual}
\end{equation}
The aggregated residual is then transferred to an unrelated target image $x_t$ to construct a forged image:
\begin{equation}
x_f
=
\operatorname{clip}_{[0,1]}
\left(
x_t+\hat{\bar{r}}_{s}
\right),
\label{eq:residual_forgery}
\end{equation}
where $\operatorname{clip}_{[0,1]}(\cdot)$ clips pixel values to the valid image intensity range. During verification, the decoder may recognize $x_f$ as carrying the same watermark as the reference images, leading to false ownership or provenance attribution. 
Existing residual-based forgery attacks mainly differ in how they reconstruct the clean counterparts or directly estimate the watermark residuals.

\subsection{Threat Model}
\paragraph{Adversarial goal.}
The adversary seeks to make an unrelated target image be verified as carrying the same watermark message as one or more released watermarked reference images. The forged image should preserve the semantic content and visual quality of the target image, while being falsely attributed to the source represented by the target watermark.
\paragraph{Adversary capabilities.}
Our primary threat model follows practical residual-based forgery attacks~\citep{souvcek2026transferable}. The adversary has access to one or a few released watermarked images carrying the same target message and may use external datasets or pretrained models to estimate transferable watermark evidence. The adversary does not know the deployed watermarking method or any protection mechanism, has no access to the parameters of the watermark encoder or decoder, and cannot query the encoder to generate additional watermarked images.
\section{Understanding Residual Transferability}
\label{sec:understand_rt}

In this section, we systematically study residual transferability in neural image watermarking systems.
First, we introduce an oracle transfer protocol that measures this property while removing the confounding effect of attack-specific residual estimation.
Second, we conduct a comparative behavioral analysis across watermarking systems exhibiting different levels of residual transferability, from both the encoder and decoder perspectives.
Guided by these observations, we then turn to systems with low residual transferability and perform targeted architectural interventions to identify the design choices underlying their low-transferability behavior.

\subsection{Measuring Residual Transferability}
\label{sec:rt_measurement}

To measure watermark residual transferability without confounding from a
particular residual-estimation attack, we use ground-truth (GT) residuals.
For each message $m_k$ from a set of $M$ messages, we construct an oracle
residual from $s$ source images:
\begin{equation}
\bar{r}_{s}^{(k)}
=
\frac{1}{s}
\sum_{i=1}^{s}
\left(
x_{r,i}^{w,(k)} - x_{r,i}^{(k)}
\right),
\end{equation}
where $x_{r,i}^{w,(k)}$ is obtained by embedding message $m_k$ into the
clean source image $x_{r,i}^{(k)}$.

We then transfer each residual $\bar{r}_{s}^{(k)}$ to a disjoint set of
$T$ target images $\{x_{t,j}\}_{j=1}^{T}$ and compute the average transfer
accuracy over both messages and target images:
\begin{equation}
\bar{a}_s
=
\frac{1}{MT}
\sum_{k=1}^{M}
\sum_{j=1}^{T}
\operatorname{BitAcc}
\left(
D\!\left(
\operatorname{clip}_{[0,1]}
\left(
x_{t,j} + \bar{r}_{s}^{(k)}
\right)
\right),
m_k
\right).
\end{equation}

Accordingly, we define residual transferability as
\begin{equation}
\mathrm{RT}@s
=
\max\left(2\bar{a}_s-1,\,0\right).
\end{equation}
where $\mathrm{RT}@s$ provides a normalized measure of residual transferability in $[0,1]$, with $0$ corresponding to random guessing and larger values indicating stronger cross-image transfer.
Using GT residuals removes the dependence on any particular residual-estimation procedure, allowing $\mathrm{RT}@s$ to more directly characterize transferability inherent to the watermarking system itself.
Unless otherwise specified, we use $s=100$ source images, $T=100$ target images, and $M=5$ messages in the following analysis. We additionally examine the effect of different source counts $s$ in Appendix~\ref{differentsourcecounts}.

\subsection{Behavioral Signatures of High Residual Transferability}
\label{sec:behavioral_signatures}

\begin{figure}[ht]
    \centering
    \includegraphics[width=\textwidth]{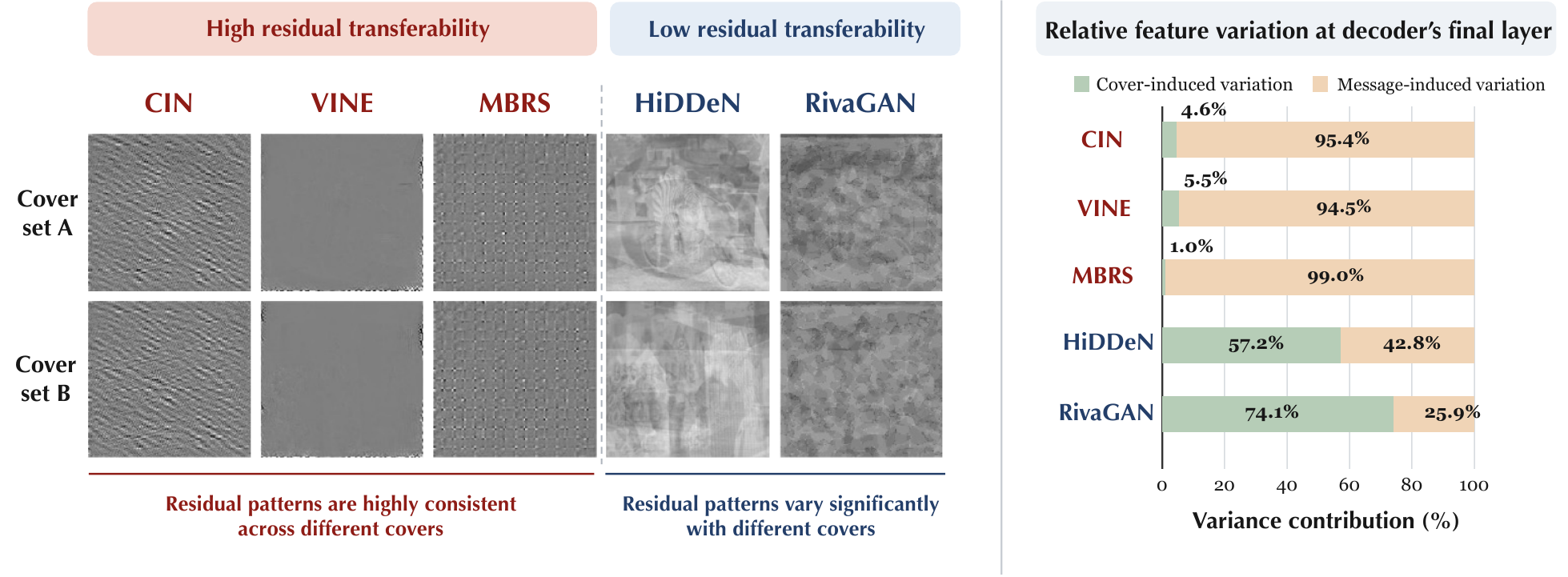}
    \caption{
        Behavioral signatures of high- and low-RT watermarking methods.
        \textbf{Left:} Average residual patterns computed from two different cover sets,
        each containing five images.
        \textbf{Right:} Relative feature variation induced by the cover image and the
        embedded message at the decoder's final layer.
    }
    \label{fig:behavioral_signatures}
\end{figure}

Before examining the underlying design factors, we first characterize how high- and low-RT watermarking systems differ in their learned behavior from two complementary perspectives: watermark embedding and watermark decoding, as summarized in Figure~\ref{fig:behavioral_signatures}.

Figure~\ref{fig:behavioral_signatures}(a) visualizes residuals averaged over groups of five cover images. Averaging suppresses cover-dependent residual components and makes structures that persist across different covers easier to observe. High-RT methods exhibit substantially more consistent residual patterns across different image groups. This behavior is consistent with the presence of a stronger cover-agnostic component in the embedded watermark signal, whereas low-RT methods exhibit residuals that vary more strongly with the underlying cover.

We next examine the representations formed during decoding. Figure~\ref{fig:behavioral_signatures}(b) compares the relative feature variation induced by the cover image and the embedded message at the decoder's final feature-extraction layer. Specifically, we measure cover-induced variation by fixing the message and varying the cover image, and message-induced variation by fixing the cover image and varying the message. The corresponding feature variances are then normalized to sum to $100\%$. High-RT decoders exhibit a substantially smaller relative contribution from cover variation, whereas low-RT systems retain a much larger cover-induced component. This contrast is consistent with a stronger dependence on the cover image during decoding in low-RT systems.

\subsection{What Prevents Residual Transferability?}
\label{sec:low_rt_designs}

The behavioral analysis above reveals a common signature of high RT, but does not explain which design choices prevent cover-agnostic watermark pathways from emerging. We therefore turn to two low-RT systems as positive cases and ask what keeps their watermark evidence dependent on the cover image. 

\begin{wrapfigure}{r}{0.48\textwidth}
    \centering
    \vspace{-15pt}
    \includegraphics[width=\linewidth]{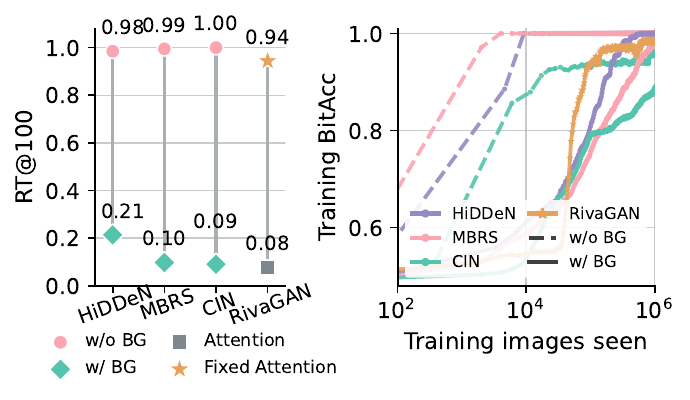}
    \caption{
    \textbf{Left:} RT under controlled architectural interventions with Broadcast--GAP (BG) and content-adaptive attention.
    \textbf{Right:} Training dynamics of high- and low-RT configurations.
    }
    \label{fig:arch_mechanism}
    \vspace{-10pt}
\end{wrapfigure}

By comparing the architectural differences across watermarking systems and ruling out common training-side factors as sufficient explanations for the RT gap (see Appendix~\ref{app:architecture_analysis}), we identify a distinctive design in HiDDeN that maintains spatial image--message interactions throughout the pipeline. During embedding of HiDDeN, the message is spatially broadcast and fused with local image features, while the decoder preserves spatial representations until a late global average pooling stage. We refer to this co-design as \emph{Broadcast--GAP (BG)}. As shown in the left panel of Figure~\ref{fig:arch_mechanism}, removing BG from HiDDeN sharply increases RT, whereas introducing this design into high-RT watermarks substantially reduces RT.\footnote{We also attempted to directly introduce BG into VINE~\citep{VINE}, but the modified model did not converge reliably given VINE's already challenging optimization.} These interventions identify BG as a design that promotes stronger dependence between watermark evidence and image content.

RivaGAN achieves low RT through a different mechanism. Its encoder contains
content-adaptive attention that modulates the embedding pattern according to the input image. Replacing this adaptive mechanism with a fixed embedding pattern causes a large increase in RT, as shown in the left panel of
Figure~\ref{fig:arch_mechanism}. This intervention identifies content
adaptivity as another mechanism that binds the watermark signal to the cover image. The complete numerical results for the architectural interventions shown in the left panel are reported in Appendix
Table~\ref{tab:cross_family_codesign_rt}.

The right panel of Figure~\ref{fig:arch_mechanism} provides complementary
evidence from training dynamics. High-RT configurations reach high training
decoding accuracy after seeing relatively few images, consistent with learning an easier cover-independent shortcut for message recovery. In contrast, configurations containing the identified low-RT mechanisms require substantially more image observations before converging. This behavior is consistent with a harder learning regime in which message recovery depends more strongly on image-conditioned representations and therefore requires exposure to a broader range of image statistics.

\section{CoverLock}
\label{sec:method}

Our analysis suggests that reducing residual transferability (RT) requires binding the embedded payload to the content of covers. Directly introducing such a dependency into an existing watermark scheme, however, typically requires model-specific redesign and retraining.
We therefore propose \textbf{CoverLock}, a watermark-agnostic plug-in that
wraps the message with a stable, image-conditioned binary code.
CoverLock is trained independently of the underlying watermarking system and can be attached to an existing system without modifying its architecture or parameters.

\begin{figure*}[ht]
    \centering
    \includegraphics[width=\textwidth]{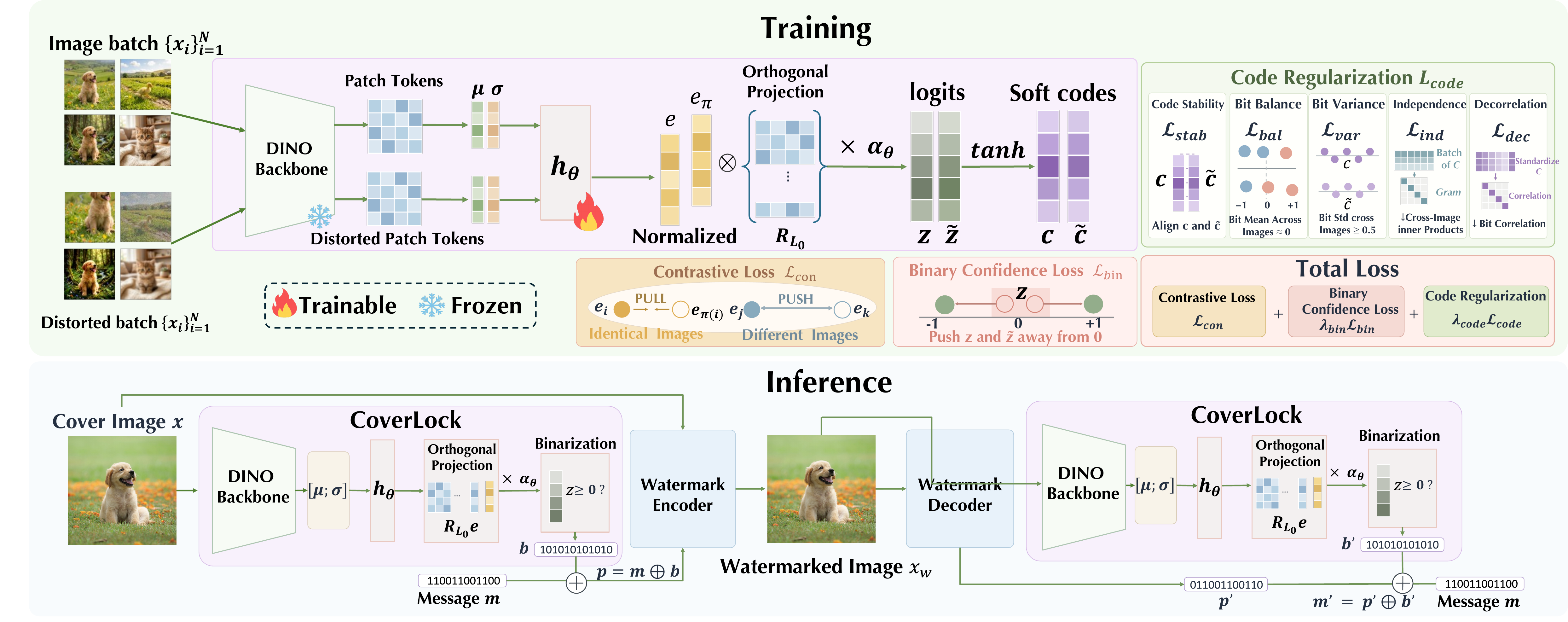}
    \caption{\textbf{Overview of CoverLock.}
    CoverLock derives a distortion-stable, image-conditioned binary code from
    frozen DINOv2 features and uses it to wrap and unwrap the watermark
    payload.}
    \label{fig:coverlock_overview}
\end{figure*}

\paragraph{Plug-in image-conditioned wrapping.}
Let $x$ denote a carrier image and
$m\in\{0,1\}^{L}$ an $L$-bit message.
CoverLock derives an image-conditioned binary code
$b_{\theta}(x)\in\{0,1\}^{L}$ and uses it to wrap the message before
watermark embedding:
\begin{equation}
    p
    =
    m\oplus b_{\theta}(x),
    \qquad
    x^{w}
    =
    E(x,p),
\end{equation}
where $E$ denotes the original watermark encoder and $\oplus$ is bitwise XOR.
At detection time, the original decoder first recovers the wrapped payload,
while CoverLock recomputes the image-conditioned code from the received image:
\begin{equation}
    \widehat{m}
    =
    D(\tau(x^{w}))
    \oplus
    b_{\theta}(\tau(x^{w})),
    \qquad
    \tau\sim\mathcal{T}.
\end{equation}
Thus, the same user message is mapped to different embedded payloads depending
on the carrier image, while the original watermark encoder and decoder remain
unchanged.

\paragraph{Image-conditioned CoverLock hash.}
To generate a carrier-specific yet distortion-stable code, we extract dense
patch features $\{q_p\}_{p=1}^{P}$ from the final layer of a frozen DINOv2~\citep{oquab2023dinov2} backbone.
We summarize the patch features using their channel-wise mean and standard deviation, which capture complementary information about the global feature response and its spatial variation across image regions:
\begin{equation}
    s(x)
    =
    \left[
        \operatorname{Mean}_{p}(q_p);
        \operatorname{Std}_{p}(q_p)
    \right].
\end{equation}
The resulting statistics are mapped by a lightweight trainable projector $h_{\theta}$ and $\ell_2$-normalized to obtain the image embedding
\begin{equation}
    e_{\theta}(x)
    =
    \operatorname{norm}_{2}
    \bigl(h_{\theta}(s(x))\bigr).
\end{equation}

We then map the embedding to the hash space using a fixed orthogonal projection.
Let $R_{L_0}\in\mathbb{R}^{D\times L_{0}}$ denote the base projection matrix, where $L_{0}=256$ and
$R_{L_0}^{\top}R_{L_0}=I$.
For a watermark system with payload length $L\leq L_{0}$, we use a fixed subset of $L$ projection directions,
$R_{L}\in\mathbb{R}^{D\times L}$.
The projected logits and corresponding soft codes are computed as
\begin{equation}
    z_{\theta}(x)
    =
    \alpha_{\theta}R_{L_0}^{\top}e_{\theta}(x),
    \qquad
    c_{\theta}(x)
    =
    \tanh\bigl(z_{\theta}(x)\bigr),
\end{equation}
where $\alpha_{\theta}$ controls the scale of the projected logits. The binary code used for message wrapping is obtained by thresholding the logits:
\begin{equation}
    b_{\theta}(x)
    =
    \mathbbm{1}
    [z_{\theta}(x)\geq 0].
\end{equation}
The fixed orthogonal projection provides distinct decision directions for different bits without introducing trainable bit-specific predictors.

\paragraph{Watermark-agnostic contrastive training.}
CoverLock is trained solely from natural images in MSCOCO~\citep{MS-COCO} and their transformed views,
without involving the watermark encoder $E$, decoder $D$, or any forgery
samples. For each training image $x_i$, we sample
$\tau_i\sim\mathcal{T}$ and construct a transformed view
$\widetilde{x}_i=\tau_i(x_i)$. The clean and transformed views of the same
image form a positive pair, while views from different images act as
negatives.
Let $\{e_j\}_{j=1}^{2N}$ denote the normalized embeddings of the $N$ clean
images and their distorted counterparts, and let $\pi(j)$ denote the paired
view of $e_j$. We optimize the symmetric contrastive objective
\begin{equation}
    \mathcal{L}_{\mathrm{con}}
    =
    -\frac{1}{2N}
    \sum_{j=1}^{2N}
    \log
    \frac{
        \exp(e_j^{\top}e_{\pi(j)}/\tau_c)
    }{
        \sum_{k\neq j}
        \exp(e_j^{\top}e_k/\tau_c)
    },
\end{equation}
where $\tau_c$ is the contrastive temperature.
To further obtain stable and informative binary codes, we introduce a
binary-confidence loss $\mathcal{L}_{\mathrm{bin}}$ and a code regularizer
$\mathcal{L}_{\mathrm{code}}$. The overall objective is
\begin{equation}
    \mathcal{L}
    =
    \mathcal{L}_{\mathrm{con}}
    +
    \lambda_{\mathrm{bin}}\mathcal{L}_{\mathrm{bin}}
    +
    \lambda_{\mathrm{code}}\mathcal{L}_{\mathrm{code}}.
\end{equation}
The transformation, individual code regularizers, and training details are provided in Appendix~\ref{app:coverlock_details}.
\section{Experiments}

\subsection{Setup}
\label{sec:exp_setup}

We evaluate CIN~\citep{CIN}, VINE~\citep{VINE}, and MBRS~\citep{jia2021mbrs} as representative high-RT schemes identified as vulnerable in prior residual-based forgery studies~\citep{souvcek2026transferable}.
We consider two defense baselines. For classifier-based filtering, we train a ConvNeXt~\citep{liu2022convnet} binary classifier for each watermark family using genuine watermarked images and GT-residual forgeries. We select ConvNeXt because it achieves the strongest detection performance among the classification backbones we evaluated (see Appendix~\ref{app:classifiers}). The second baseline follows MHDW~\citep{lu2006mhdw}, which, similar in spirit to our approach, binds watermark verification to image content through a robust media hash. In contrast, CoverLock constructs its image-dependent code through objectives jointly optimized for security and robustness.

\paragraph{Datasets.}
We evaluate all methods on three image datasets: DIV2K~\citep{div2k}, ImageNet~\citep{Imagenet}, and MS-COCO~\citep{MS-COCO}. These datasets have diverse image distributions and resolutions, allowing us to evaluate residual-based forgery across different visual domains. For each dataset, images used for residual estimation and those used as forgery targets are randomly sampled from two disjoint subsets. 

\paragraph{Residual-based Forgery Attacks.}
We evaluate security against residual-based forgery attacks, where an attacker estimates the watermark residual from one or more released watermarked images and transfers it to an unrelated target image. We consider both single- and multi-reference settings: GT@1 uses one ground-truth residual,~\citet{yang2024steganalysis} uses 100 watermarked references, and WMForger~\citep{souvcek2026transferable} uses a single watermarked reference, following their original configurations unless otherwise specified.

\paragraph{Metrics.}
We report True Positive Rate (TPR) and Attack Success Rate (ASR) at the sample level, and Bit Accuracy (BitAcc) and Bit Error Rate (BER) at the bit level. TPR measures the fraction of legitimate watermarked images that are accepted, while ASR corresponds to the false-positive rate on forged images. BitAcc and BER measure the fractions of correctly and incorrectly decoded watermark bits, respectively. Verification thresholds are calibrated under the Bernoulli null model to FPR $<10^{-6}$ for all watermark schemes. Higher TPR and BitAcc, and lower ASR and BER, are better.

\subsection{Main Results of CoverLock}

We first evaluate whether CoverLock can prevent residual-based watermark
forgery across different watermarking systems, image domains, and attack methods.

Table~\ref{tab:Coverlock-security} reports the ASR under GT@1,
\citet{yang2024steganalysis}, and WmForger~\citep{souvcek2026transferable}.
CoverLock consistently achieves low ASR across all watermarking systems,
datasets, and attacks, with even its highest average ASR remaining below
0.25\%. In contrast, the classifier-based baseline generalizes poorly across attacks, as the residual artifacts learned from GT-based forgeries can change substantially under different estimation procedures. This dependence on known attack patterns may lead to an iterative attacker--defender cycle~\citep{li2026badwam,li2026spongetoolattackstealthy}. MHDW~\citep{lu2006mhdw} also exhibits attack-dependent protection and becomes less effective when the attacker obtains cleaner residuals through stronger averaging or more accurate estimation. By making watermark verification explicitly dependent on image content, CoverLock avoids relying on attack-specific forgery signatures and provides more consistent protection across attack methods.

\begin{table*}[t]
\centering
\caption{
Forgery attack success rate (ASR, \%) across evaluation datasets.
Results are reported as mean {\scriptsize $\pm$ std} over three random seeds.
}
\label{tab:Coverlock-security}

\small
\setlength{\tabcolsep}{1.5pt}
\renewcommand{\arraystretch}{1.2}
\resizebox{\linewidth}{!}{%
\begin{tabular}{@{}llccccccccc@{}}
\toprule
& &
\multicolumn{3}{c}{GT@1 $\downarrow$} &
\multicolumn{3}{c}{\citet{yang2024steganalysis} $\downarrow$} &
\multicolumn{3}{c}{WmForger~\citep{souvcek2026transferable} $\downarrow$} \\
\cmidrule(lr){3-5}
\cmidrule(lr){6-8}
\cmidrule(lr){9-11}

Watermark & Dataset
& ConvNeXt & MHDW & CoverLock
& ConvNeXt & MHDW & CoverLock
& ConvNeXt & MHDW & CoverLock \\
\midrule

\multirow{3}{*}{CIN}
& COCO
& \attackcell{24.12}{3.7}
& \attackcell{0.01}{0.0}
& \viscell{\textbf{0.00}}{0.0}
& \attackcell{36.16}{2.6}
& \attackcell{2.17}{0.6}
& \viscell{\textbf{0.00}}{0.0}
& \attackcell{86.09}{2.0}
& \attackcell{0.00}{0.0}
& \viscell{\textbf{0.00}}{0.0} \\

& ImageNet
& \attackcell{29.36}{3.9}
& \attackcell{\textbf{0.00}}{0.0}
& \attackcell{0.03}{0.0}
& \attackcell{36.37}{1.7}
& \attackcell{3.19}{0.9}
& \viscell{\textbf{0.00}}{0.0}
& \attackcell{86.93}{1.7}
& \attackcell{\textbf{0.00}}{0.0}
& \viscell{0.00}{0.0} \\

& DIV2K
& \attackcell{18.72}{4.6}
& \attackcell{\textbf{0.00}}{0.0}
& \viscell{\textbf{0.00}}{0.0}
& \attackcell{29.94}{1.9}
& \attackcell{4.72}{0.6}
& \viscell{\textbf{0.11}}{0.2}
& \attackcell{78.56}{2.1}
& \attackcell{\textbf{0.00}}{0.0}
& \viscell{\textbf{0.00}}{0.0} \\

\midrule

\multirow{3}{*}{MBRS}
& COCO
& \attackcell{2.41}{1.4}
& \attackcell{11.37}{4.4}
& \viscell{\textbf{0.11}}{0.0}
& \attackcell{2.37}{0.1}
& \attackcell{46.14}{2.4}
& \viscell{\textbf{0.00}}{0.0}
& \attackcell{96.28}{0.1}
& \attackcell{4.76}{1.7}
& \viscell{\textbf{0.06}}{0.0} \\

& ImageNet
& \attackcell{1.01}{0.6}
& \attackcell{10.81}{5.0}
& \viscell{\textbf{0.82}}{0.2}
& \attackcell{2.96}{0.6}
& \attackcell{49.57}{0.8}
& \viscell{\textbf{0.00}}{0.0}
& \attackcell{91.78}{1.3}
& \attackcell{3.68}{1.6}
& \viscell{\textbf{0.17}}{0.1} \\

& DIV2K
& \attackcell{\textbf{0.83}}{0.7}
& \attackcell{17.33}{9.7}
& \viscell{\textbf{0.83}}{0.1}
& \attackcell{2.44}{0.2}
& \attackcell{55.11}{0.9}
& \viscell{\textbf{0.00}}{0.0}
& \attackcell{87.17}{2.3}
& \attackcell{5.44}{4.2}
& \viscell{\textbf{0.28}}{0.2} \\

\midrule

\multirow{3}{*}{VINE}
& COCO
& \attackcell{14.44}{4.7}
& \attackcell{1.16}{0.4}
& \viscell{\textbf{0.00}}{0.0}
& \attackcell{4.39}{0.4}
& \attackcell{28.02}{1.6}
& \viscell{\textbf{0.00}}{0.0}
& \attackcell{22.34}{4.1}
& \attackcell{0.06}{0.1}
& \viscell{\textbf{0.00}}{0.0} \\

& ImageNet
& \attackcell{14.52}{4.2}
& \attackcell{0.49}{0.3}
& \attackcell{\textbf{0.11}}{0.1}
& \attackcell{9.32}{0.9}
& \attackcell{29.22}{1.0}
& \viscell{\textbf{0.02}}{0.0}
& \attackcell{25.66}{2.7}
& \attackcell{0.06}{0.1}
& \viscell{\textbf{0.01}}{0.0} \\

& DIV2K
& \attackcell{4.61}{1.2}
& \attackcell{1.56}{1.0}
& \viscell{\textbf{0.28}}{0.1}
& \attackcell{4.00}{0.4}
& \attackcell{23.06}{4.4}
& \viscell{\textbf{0.00}}{0.0}
& \attackcell{13.61}{2.8}
& \attackcell{0.06}{0.1}
& \viscell{\textbf{0.00}}{0.0} \\

\midrule

\multicolumn{2}{l}{\textbf{Average}}
& 12.226
& 4.747
& \cellcolor{visgreenstrong}\textbf{0.243}
& 14.216
& 26.800
& \cellcolor{visgreenstrong}\textbf{0.015}
& 65.379
& 1.560
& \cellcolor{visgreenstrong}\textbf{0.059} \\

\bottomrule
\end{tabular}
}
\end{table*}

\subsection{Analysis of CoverLock's Security and Robustness}
We further analyze the behavior of different forgery defenses and the security--robustness trade-off of CoverLock.

\begin{wrapfigure}{r}{0.55\textwidth}
    \centering
    \vspace{-10pt}
    \begin{minipage}{0.98\linewidth}
        \centering
    
        \includegraphics[width=\linewidth]
        {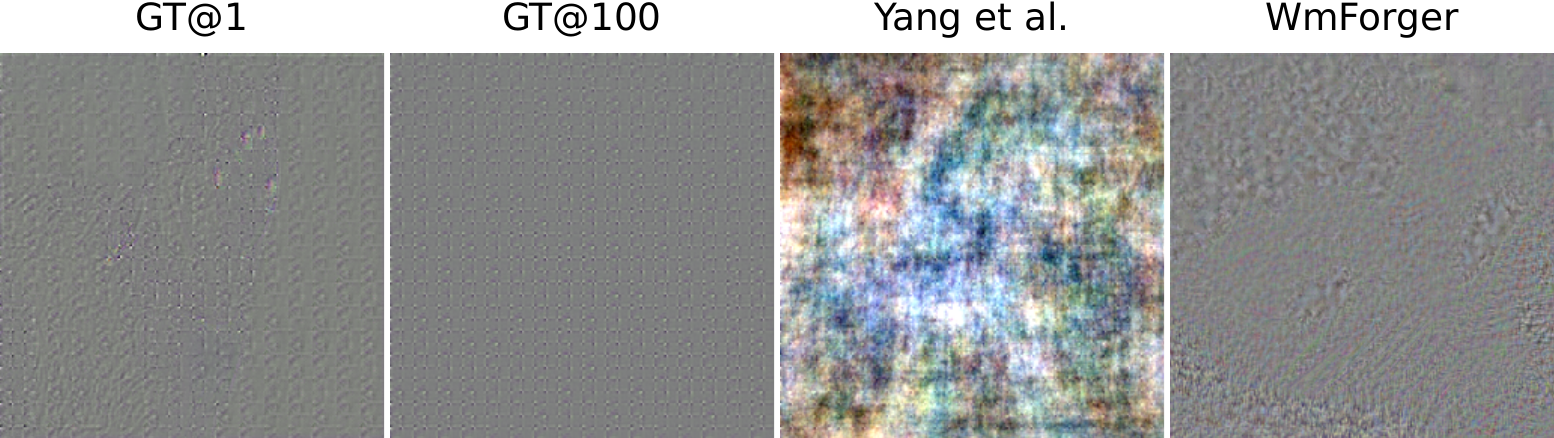}
    
        \vspace{-2pt}
        {\fontsize{7}{8}\selectfont
        (a) Residual Patterns across Attacks}
    \end{minipage}

    \vspace{4pt}
    
    \begin{minipage}[t]{0.47\linewidth}
        \centering
        \includegraphics[width=\linewidth]
        {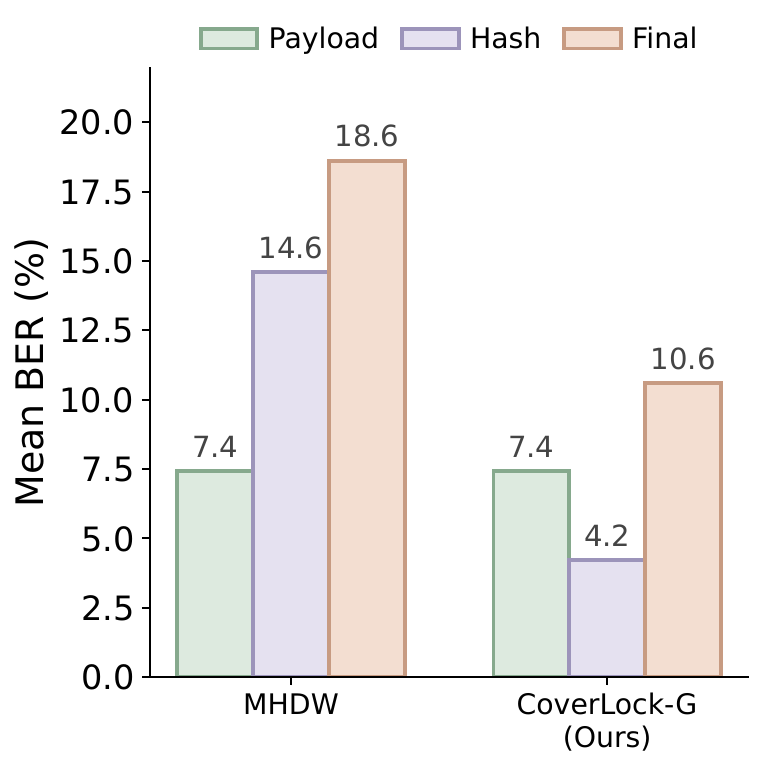}\\[-2pt]
        {\fontsize{7}{8}\selectfont
        (b) BER Decomposition}
    \end{minipage}%
    \hfill
    \begin{minipage}[t]{0.51\linewidth}
        \centering
        \includegraphics[width=\linewidth]
        {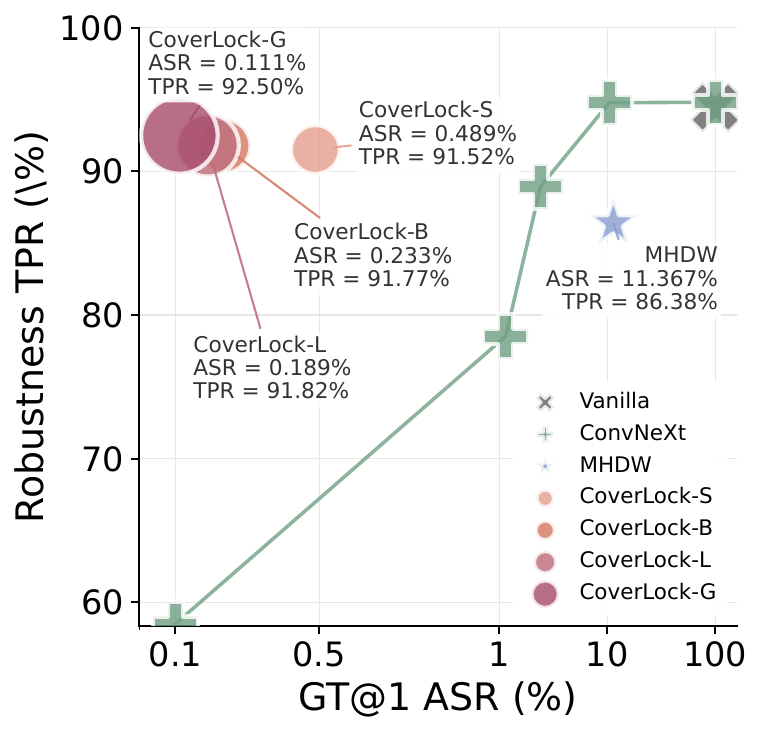}\\[-2pt]
        {\fontsize{7}{8}\selectfont
        (c) Security--Robustness}
    \end{minipage}%

    \caption{Attacks and CoverLock analysis.}
    \label{fig:coverlock_analysis}
    \vspace{-12pt}
\end{wrapfigure}

Figure~\ref{fig:coverlock_analysis}(a) visualizes the residuals produced by different forgery procedures. Even for the same watermarking system, the residual patterns vary substantially with the estimation procedure and the
number of source images. In particular, GT@1 and GT@100 exhibit different estimation quality, while~\citet{yang2024steganalysis} and WmForger~\citep{souvcek2026transferable} produce
visually distinct residual structures. This diversity helps explain the limited cross-attack generalization of classifier-based defenses in Table~\ref{tab:Coverlock-security}, which rely on artifacts observed from known forgery patterns.
Figure~\ref{fig:coverlock_analysis}(b) decomposes the mean BER under distortions into payload and image-dependent hash components, while Table~\ref{tab:binding_robustness_bitacc} reports the corresponding bit accuracy under each individual distortion. MHDW and CoverLock-G have the same payload BER of $7.4\%$, whereas CoverLock-G reduces the hash BER from $14.6\%$ to $4.2\%$, resulting in a substantially lower final BER. This suggests that CoverLock's robustness advantage mainly comes from more reliable recovery of the image-dependent component rather than improved payload decoding.
Figure~\ref{fig:coverlock_analysis}(c) compares the resulting security--robustness trade-off for MBRS on COCO. For the classifier baseline, reducing ASR requires a substantial loss in TPR, while operating points with high TPR remain considerably more vulnerable to forgery. MHDW similarly provides a less favorable trade-off. In contrast, CoverLock variants remain concentrated in the low-ASR, high-TPR region. Increasing the DINOv2 backbone size from Small to Giant further reduces ASR while maintaining TPR at approximately $92\%$, with CoverLock-G lying on the favorable Pareto frontier.

\begin{table*}[t]
\centering
\caption{Bit accuracy (BitAcc, \%) on COCO under Gaussian noise ($\sigma=0.05$), brightness ($\times 1.2$), Gaussian blur ($\sigma=1.0$), JPEG compression ($Q=70$), contrast ($\times 1.2$), and center crop (80\% area). Average is over the six distortions and bold indicates the best binding defense, excluding Vanilla.}
\label{tab:binding_robustness_bitacc}
\small
\setlength{\tabcolsep}{4.2pt}
\renewcommand{\arraystretch}{0.96}
\begin{tabular}{@{}llccccccc@{}}
\toprule
\textbf{Watermark} & \textbf{Method} & \textbf{Noise} & \textbf{Brightness} & \textbf{Blur} & \textbf{JPEG} & \textbf{Contrast} & \textbf{Crop} & \textbf{Average} \\
\midrule
\rowcolor{rowgray}
CIN & Vanilla & 99.98 & 99.92 & 100.00 & 96.28 & 100.00 & 100.00 & 99.36 \\
 & MHDW & 76.49 & 91.08 & 79.83 & 82.19 & 91.74 & 83.20 & 84.09 \\
 & CoverLock & \textbf{93.35} & \textbf{97.13} & \textbf{95.58} & \textbf{91.52} & \textbf{97.45} & \textbf{92.55} & \textbf{94.60} \\
\midrule
\rowcolor{rowgray}
MBRS & Vanilla & 69.16 & 98.52 & 94.13 & 100.00 & 99.06 & 94.41 & 92.55 \\
 & MHDW & 59.92 & 88.57 & 83.81 & 90.30 & 89.43 & 76.32 & 81.39 \\
 & CoverLock & \textbf{66.43} & \textbf{96.16} & \textbf{91.62} & \textbf{96.65} & \textbf{96.75} & \textbf{88.82} & \textbf{89.40} \\
\midrule
\rowcolor{rowgray}
VINE & Vanilla & 99.72 & 99.85 & 99.94 & 99.91 & 99.96 & 50.27 & 91.61 \\
 & MHDW & 82.27 & 93.11 & 88.04 & 91.62 & 93.50 & 50.20 & 83.12 \\
 & CoverLock & \textbf{94.08} & \textbf{95.90} & \textbf{95.39} & \textbf{94.89} & \textbf{95.98} & \textbf{50.25} & \textbf{87.75} \\
\bottomrule
\end{tabular}
\end{table*}

\section{Related work}\label{sec:relatedwork}

\paragraph{Image Watermarking}
Existing image watermarking methods can be broadly categorized into \emph{post-processing} and \emph{in-generation} approaches. 
Post-processing methods embed watermarks into already generated images. Early approaches based on handcrafted embedding (e.g., LSB~\citep{chopra2012lsb}, DWT-DCT~\citep{DWT-DCT,DWT-DCT-SVD}) demonstrate feasibility but are vulnerable to common image transformations. Recent neural approaches improve robustness via learned encoder--decoder architectures with adversarial objectives~\citep{zhu2018hidden,jia2021mbrs,stegastamp,RivaGAN,CIN,PIMoG,VINE,sander2025watermark,soucek2025pixelseal}. These methods are compatible with existing generative systems without architectural modifications.
In contrast, in-generation watermarking incorporates watermark signals directly during image synthesis rather than embedding them into already generated images~\citep{Gaussianshading,wen2023tree,stablesignature,arabi2025seal,wang2025sleepermark,chen2025tagwm,lee2025semanticwatermarking,liu2025harnessingfrequencyspectruminsights}.
In this work, we focus on post-processing methods due to their model-agnostic and deployment-friendly properties.

\paragraph{Forgery Attack and Defense.}
Watermark forgery attacks estimate transferable watermark evidence from one or a few watermarked images and apply it to unrelated covers~\citep{kutter2000watermark,yang2024steganalysis,souvcek2026transferable}. 
\citet{dong2025wmcopier} instead assumes access to a large collection of watermarked images with the same message, which is outside our few-shot threat model. 
Prior defenses either bind watermarks to image content~\citep{lu2006mhdw} or improve forgery resistance for in-generation watermarking by intervening in the diffusion process, e.g., conditioning latent watermarks on image semantics, modifying the denoising trajectory, or constraining intermediate latents~\citep{zhang2026sembind,liu2026pai,lai2026csguard}. 
In contrast, we focus on strengthening cover dependence for post-processing neural watermarking systems.

\section{Conclusion}

In this work, we study watermark forgery through \emph{residual transferability} (RT), revealing that many deep watermarking systems can develop cover-agnostic watermark pathways whose signals transfer across unrelated images. By analyzing low-RT systems, we further identify coupled spatial interaction and content-adaptive embedding as two architectural mechanisms associated with stronger cover dependence.
We also introduce CoverLock, a complementary solution for existing systems without redesigning or retraining the underlying watermark model. Across different watermarking systems, datasets, and attacks, CoverLock achieves a more favorable balance between forgery resistance and robustness than prior media-hash-based defenses, further highlighting cover dependence as an important design principle for secure deep image watermarking.

\section*{AI Use Statement}
LLM tools were used to assist with manuscript writing and editing, as well as with minor modifications to Python utility scripts, such as plotting and visualization scripts. All generated content and code modifications were reviewed and verified by the authors, who take full responsibility for the final content of this work.

\section*{Ethics Statement}
This work analyzes watermark forgery attacks introduced in prior work, with the goal of understanding their underlying mechanisms and developing effective defenses. Our study focuses on security analysis and on deriving design insights for more forgery-resistant image watermarking systems.

\section*{Reproducibility Statement}
We provide detailed experimental settings, model configurations,
training procedures, evaluation protocols, and additional ablation results
in the appendices. The datasets and pretrained watermarking models used in
our evaluation are publicly available. We will release the implementation
of our CoverLock to facilitate reproduction of the reported results.

\bibliography{iclr2027_conference}
\bibliographystyle{iclr2027_conference}
\appendix
\section{Limitation}
Although CoverLock achieves low forgery success and a favorable balance between forgery resistance and robustness under the distortions considered
in our evaluation, this balance may change under more severe or unseen image transformations. Its behavior under broader distribution shifts and
adaptive attacks specifically targeting the image-binding mechanism also remains to be studied.
CoverLock provides a lightweight way to augment existing watermarking systems without redesigning their core architectures, but introduces an
additional image-binding module and associated training and deployment overhead. Finally, while our analysis covers several representative neural
watermarking architectures and identifies two mechanisms associated with low RT, other architectures may achieve cover dependence through different design principles.

\section{CoverLock Implementation and Training Details}
\label{app:coverlock_details}

\subsection{Architecture and Orthogonal Hashing}

\paragraph{Frozen feature extractor.}
CoverLock uses a frozen DINOv2 ViT~\citep{oquab2023dinov2} as the
image feature extractor. Unless otherwise specified, we use ViT-g/14; the CoverLock-S/B/L/g variants replace it with the corresponding DINOv2 Small, Base, Large, and Giant backbones while leaving the
remaining architecture unchanged. All images are resized to $224\times224$, and patch tokens are extracted from the final transformer layer. This yields a $16\times16$ grid of $P=256$ patch
features,
\begin{equation}
    Q(x)
    =
    \{q_p\}_{p=1}^{P},
    \qquad
    q_p\in\mathbb{R}^{d},
\end{equation}
where $d$ depends on the selected backbone ($d=1536$ for ViT-g/14).

We summarize the patch features using their channel-wise mean and standard deviation across spatial tokens
\begin{equation}
    \mu(x)
    =
    \frac{1}{P}\sum_{p=1}^{P}q_p,
    \qquad
    \sigma(x)
    =
    \operatorname{Std}_{p}[q_p],
\end{equation}
and concatenate them to form
\begin{equation}
    s(x)
    =
    [\mu(x);\sigma(x)]
    \in\mathbb{R}^{2d}.
\end{equation}

\paragraph{Projection head.}
The pooled representation is mapped to a $D$-dimensional image embedding
using a lightweight trainable projection head,
\begin{equation}
    h_\theta(s(x))
    =
    W_2
    \operatorname{GELU}
    \left(
        W_1\operatorname{LN}(s(x))+b_1
    \right)
    +b_2,
\end{equation}
followed by $\ell_2$ normalization,
\begin{equation}
    e_\theta(x)
    =
    \frac{h_\theta(s(x))}
    {\|h_\theta(s(x))\|_2}.
\end{equation}
We use a hidden dimension of 512 and an output embedding dimension
$D=512$.

\paragraph{Fixed orthogonal projection.}
CoverLock maintains a native code length of $L_0=256$.
We first sample a Gaussian matrix
\begin{equation}
    G\in\mathbb{R}^{D\times L_0}
\end{equation}
using a fixed seed and orthogonalize its columns once using QR
decomposition. This produces a fixed projection matrix
\begin{equation}
    R_{L_0}\in\mathbb{R}^{D\times L_0},
    \qquad
    R_{L_0}^{\top}R_{L_0}=I,
\end{equation}
which is stored as a non-trainable buffer.

For a watermarking system with payload length $L\leq L_0$, a
seed-determined fixed permutation selects $L$ columns from
$R_{L_0}$,
\begin{equation}
    R_{L}
    =
    [r_{\pi(1)},\ldots,r_{\pi(L)}]
    \in\mathbb{R}^{D\times L}.
\end{equation}
The resulting hash logits are
\begin{equation}
    z_\theta^{(L_0)}(x)
    =
    \alpha_\theta R_{L}^\top e_\theta(x),
\end{equation}
where $\alpha_\theta>0$ is a learned logit scale that controls the magnitude of the projected responses. Since the image embedding and projection directions are normalized, this scale allows the model to adapt the confidence of the continuous hash logits without changing their signs, facilitating the subsequent soft-code regularization and binary-margin objective.
The corresponding binary image-dependent code is obtained by thresholding the logits,
\begin{equation}
    b_\theta^{(L)}(x)
    =
    \mathbbm{1}
    \left[
        z_\theta^{(L)}(x)\geq0
    \right].
\end{equation}

\subsection{Code Regularization and Binary Confidence}
\label{app:code_regularization}

The contrastive objective introduced in the main paper structures the
continuous image embedding $e_\theta(x)$ by bringing different views of
the same carrier together while separating different carriers.
In addition, we regularize the projected codes to encourage distortion
stability, balanced and diverse bit usage, cross-carrier independence,
and low inter-bit correlation. A separate binary-confidence objective
further encourages the projected logits to remain away from the
binarization threshold.

For a minibatch of $N$ clean images
$\{x_i\}_{i=1}^{N}$, we independently sample
$\tau_i\sim\mathcal{T}$ and construct distorted views
\begin{equation}
    \widetilde{x}_i
    =
    \tau_i(x_i).
\end{equation}
During CoverLock training, the code-level objectives are applied to
the native $L_0$-dimensional projection. We therefore define
\begin{equation}
    z_i
    =
    z_\theta^{(L_0)}(x_i),
    \qquad
    \widetilde{z}_i
    =
    z_\theta^{(L_0)}(\widetilde{x}_i),
\end{equation}
with the corresponding soft codes
\begin{equation}
    c_i
    =
    \tanh(z_i),
    \qquad
    \widetilde{c}_i
    =
    \tanh(\widetilde{z}_i).
\end{equation}

\paragraph{Code stability.}
Different views of the same carrier should produce consistent codes.
We therefore minimize
\begin{equation}
    \mathcal{L}_{\mathrm{stab}}
    =
    \frac{1}{NL_0}
    \sum_{i=1}^{N}
    \left\|
        c_i-\widetilde{c}_i
    \right\|_2^2.
\end{equation}

\paragraph{Bit balance.}
To prevent individual bits from being biased toward one binary value
over the training distribution, we encourage each bit to have zero
mean across carriers:
\begin{equation}
    \mathcal{L}_{\mathrm{bal}}
    =
    \frac{1}{L_0}
    \left\|
        \frac{1}{N}
        \sum_{i=1}^{N}c_i
    \right\|_2^2.
\end{equation}

\paragraph{Bit variance.}
Bit balance alone does not prevent a bit from remaining close to zero
for all carriers. We therefore encourage each bit to maintain
sufficient variation across images:
\begin{equation}
    \mathcal{L}_{\mathrm{var}}
    =
    \frac{1}{2L_0}
    \sum_{b=1}^{L_0}
    \left[
        \max
        \left(
            0,\nu-\operatorname{Std}_{i}[c_{ib}]
        \right)
        +
        \max
        \left(
            0,\nu-\operatorname{Std}_{i}[\widetilde{c}_{ib}]
        \right)
    \right],
\end{equation}
where $\nu=0.5$.

\paragraph{Cross-carrier independence.}
Codes from different carriers should remain dissimilar. Let
\begin{equation}
    C
    =
    [c_1^\top;\ldots;c_N^\top]
    \in\mathbb{R}^{N\times L_0}.
\end{equation}
We penalize pairwise similarities between different carriers:
\begin{equation}
    \mathcal{L}_{\mathrm{ind}}
    =
    \operatorname{mean}
    \left[
        \operatorname{offdiag}
        \left(
            \frac{CC^\top}{L_0}
        \right)^2
    \right].
\end{equation}

\paragraph{Bit decorrelation.}
Orthogonal projection directions do not necessarily guarantee
uncorrelated outputs over the image distribution. We therefore
standardize each bit over the minibatch,
\begin{equation}
    \overline{c}_{ib}
    =
    \frac{
        c_{ib}-\mu_b
    }{
        \sigma_b+\epsilon
    },
\end{equation}
where $\mu_b$ and $\sigma_b$ are the batch mean and standard deviation of bit $b$, and $\epsilon=10^{-4}$ is used for numerical stability. Let $\overline{C}\in\mathbb{R}^{N\times L_0}$ denote the resulting
standardized code matrix. We minimize its off-diagonal bit
correlations:
\begin{equation}
    \mathcal{L}_{\mathrm{dec}}
    =
    \operatorname{mean}
    \left[
        \operatorname{offdiag}
        \left(
            \frac{
                \overline{C}^{\top}\overline{C}
            }{N}
        \right)^2
    \right].
\end{equation}

The complete code regularization objective is
\begin{equation}
\begin{aligned}
    \mathcal{L}_{\mathrm{code}}
    ={}&
    \lambda_{\mathrm{stab}}\mathcal{L}_{\mathrm{stab}}
    +
    \lambda_{\mathrm{bal}}\mathcal{L}_{\mathrm{bal}}
    +
    \lambda_{\mathrm{var}}\mathcal{L}_{\mathrm{var}}
    \\[-1mm]
    &+
    \lambda_{\mathrm{ind}}\mathcal{L}_{\mathrm{ind}}
    +
    \lambda_{\mathrm{dec}}\mathcal{L}_{\mathrm{dec}}.
\end{aligned}
\end{equation}

\paragraph{Binary confidence.}
Since the final image-dependent code is obtained by thresholding the
projected logits at zero, logits close to the decision boundary are
particularly susceptible to bit flips under image distortions. We
therefore encourage both clean and distorted logits to remain away
from zero:
\begin{equation}
    \mathcal{L}_{\mathrm{bin}}
    =
    \frac{1}{2NL_0}
    \sum_{i=1}^{N}\sum_{b=1}^{L_0}
    \left[
        \operatorname{softplus}
        \left(
            \rho-|z_{ib}|
        \right)
        +
        \operatorname{softplus}
        \left(
            \rho-|\widetilde{z}_{ib}|
        \right)
    \right],
\end{equation}
where $\rho=1$ denotes the binary margin.

We use
\begin{equation}
\begin{aligned}
    \lambda_{\mathrm{stab}} &= 2,
    &\lambda_{\mathrm{bal}} &= 5,
    &\lambda_{\mathrm{var}} &= 5,\\
    \lambda_{\mathrm{ind}} &= 1,
    &\lambda_{\mathrm{dec}} &= 0.05,
    &\lambda_{\mathrm{bin}} &= 0.0125,\\
    \lambda_{\mathrm{code}} &= 0.25.
\end{aligned}
\end{equation}

Together with the contrastive objective defined in the main paper,
the complete CoverLock training objective is
\begin{equation}
    \mathcal{L}
    =
    \mathcal{L}_{\mathrm{con}}
    +
    \lambda_{\mathrm{bin}}
    \mathcal{L}_{\mathrm{bin}}
    +
    \lambda_{\mathrm{code}}
    \mathcal{L}_{\mathrm{code}}.
\end{equation}

\subsection{Training Configuration}
\label{app:coverlock_training}

CoverLock is trained solely from natural images and their distorted versions. For each image $x_i$, the second view $\widetilde{x}_i$ is generated by sampling a transformation from the training distortion distribution $\mathcal{T}$, including JPEG compression (quality 45--90), Gaussian blur ($\sigma=0.3$--$2.0$), Gaussian noise ($\sigma=0.005$--$0.02$), Crop that retains 70\%--95\% of the image area, and brightness or contrast adjustment with factors in $[0.7,1.3]$.
We optimize only the lightweight projection head and the learnable logit scale using AdamW~\citep{loshchilov2019decoupled} with a learning rate of $3\times10^{-5}$. The DINOv2 backbone and orthogonal projection matrix remain fixed. We use a batch size of 256, contrastive temperature $\tau_c=0.1$, and train the models reported in the main paper for 200k optimization steps. All CoverLock variants are trained on COCO train2017, which contains 118,287 images. We use 94,629 images (80\%) for training and reserve the
remaining 23,658 images for validation.

\subsection{Training Dynamics of CoverLock}
\label{sec:coverlock_training}

\begin{figure*}[ht]
    \centering

    \begin{subfigure}[t]{0.3\textwidth}
        \centering
        \includegraphics[width=\linewidth]
        {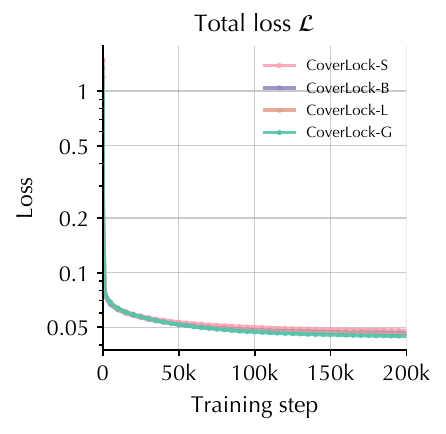}
    \end{subfigure}
    \hfill
    \begin{subfigure}[t]{0.3\textwidth}
        \centering
        \includegraphics[width=\linewidth]
        {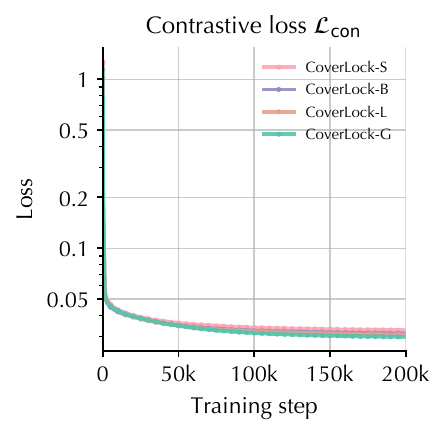}
    \end{subfigure}
    \hfill
    \begin{subfigure}[t]{0.3\textwidth}
        \centering
        \includegraphics[width=\linewidth]
        {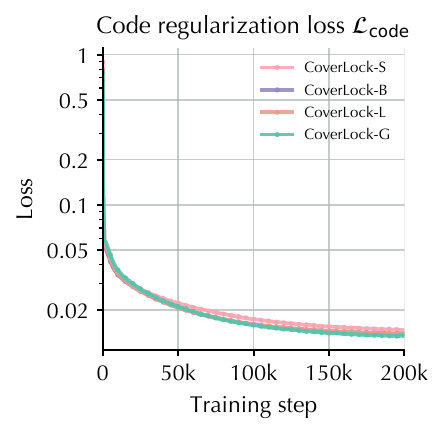}
    \end{subfigure}

    \vspace{0.4em}

    \begin{subfigure}[t]{0.3\textwidth}
        \centering
        \includegraphics[width=\linewidth]
        {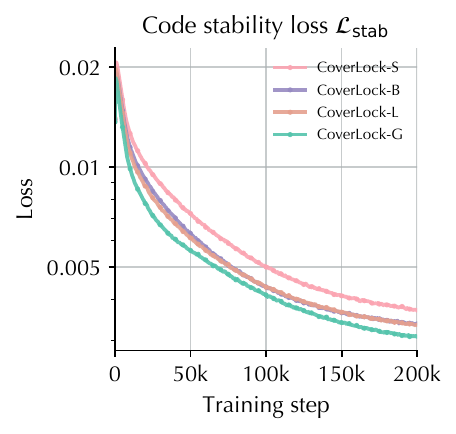}
    \end{subfigure}
    \hfill
    \begin{subfigure}[t]{0.3\textwidth}
        \centering
        \includegraphics[width=\linewidth]
        {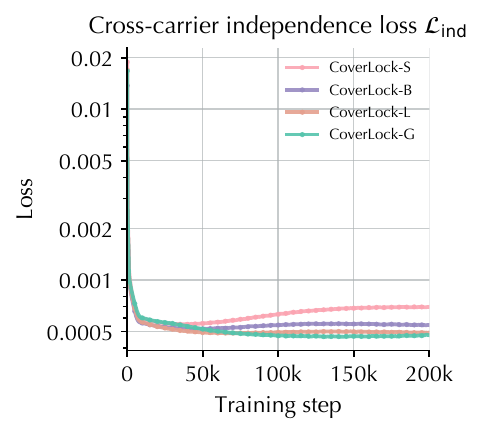}
    \end{subfigure}
    \hfill
    \begin{subfigure}[t]{0.3\textwidth}
        \centering
        \includegraphics[width=\linewidth]
        {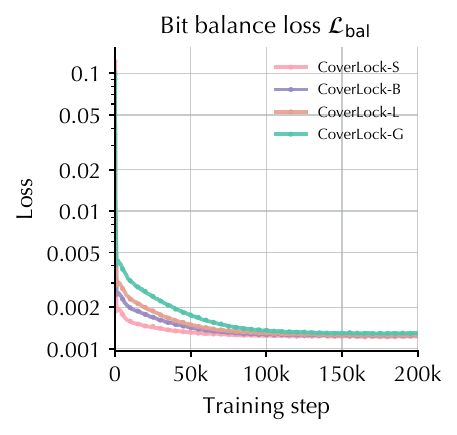}
    \end{subfigure}

    \vspace{0.4em}

    \begin{subfigure}[t]{0.3\textwidth}
        \centering
        \includegraphics[width=\linewidth]
        {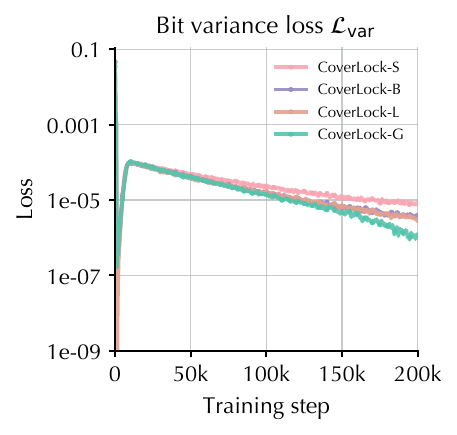}
    \end{subfigure}
    \hfill
    \begin{subfigure}[t]{0.3\textwidth}
        \centering
        \includegraphics[width=\linewidth]
        {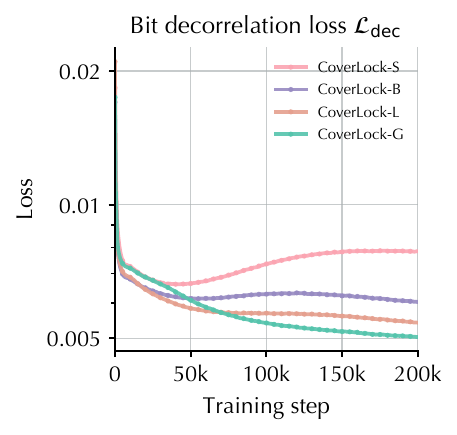}
    \end{subfigure}
    \hfill
    \begin{subfigure}[t]{0.3\textwidth}
        \centering
        \includegraphics[width=\linewidth]
        {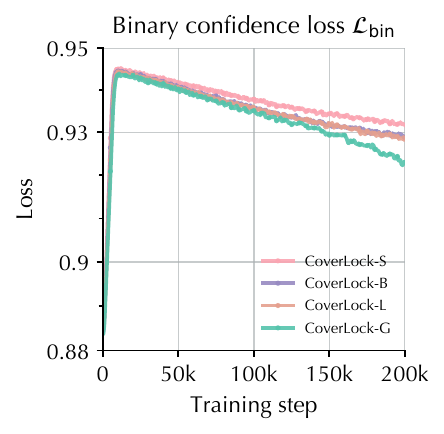}
    \end{subfigure}

    \caption{
        \textbf{Training dynamics of CoverLock.}
        Training curves of the overall objective and its individual
        loss components across four CoverLock variants.
        While the components exhibit different optimization trajectories,
        they eventually reach stable regimes under joint optimization.
    }
    \label{fig:coverlock_training}
\end{figure*}

Figure~\ref{fig:coverlock_training} shows the optimization trajectories of the overall CoverLock objective and its individual components. The total, contrastive, and code-regularization losses decrease rapidly early in training and then improve more gradually, while the stability and balance objectives also decrease steadily.
The independence and decorrelation losses exhibit a backbone-dependent late-stage rebound after their initial reduction; this rebound becomes less pronounced as the backbone size increases and is largely absent for CoverLock-G, suggesting that higher-capacity features may better accommodate
these statistical code constraints during joint optimization.
The variance and binary-confidence terms follow distinct non-monotonic trajectories as the projected codes evolve during training. Despite these different dynamics, all four variants remain well-behaved throughout optimization.

\section{Apparent Signs of Image Dependence Can Be Misleading}

As shown in Section~\ref{sec:low_rt_designs}, explicitly combining image and message features does not necessarily prevent a cover-agnostic watermark pathway from emerging. Similarly, an image-dependent-looking residual does not necessarily indicate true image dependence. A single residual can retain visible structures of the host image simply due to image-specific reconstruction errors introduced by the encoder, rather than watermark evidence used for decoding.

\begin{figure}[t]
    \centering
    \includegraphics[width=\linewidth]{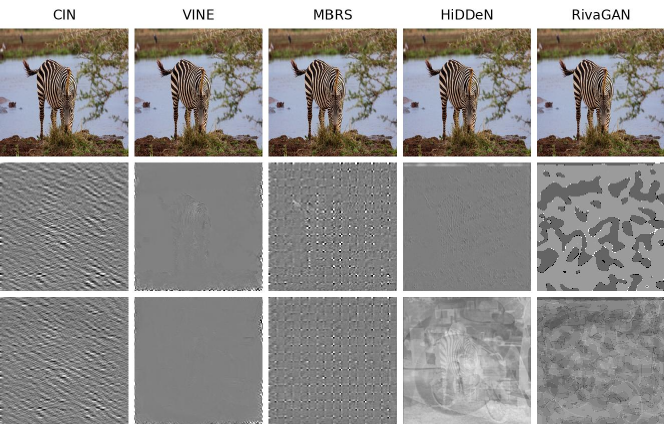}
    \caption{
    \textbf{Comparison of single-image and averaged watermark residuals.}
    From top to bottom, we show the watermarked image, the residual from a single image, and the residual averaged over five images carrying the same message for different watermarking methods.
    }
    \label{fig:cover_single_mean_3x5_grid}
\end{figure}

This motivates the residual averaging used in the main paper. Averaging residuals from images carrying the same message suppresses image-specific reconstruction errors while retaining watermark components shared across covers. As shown in Figure~\ref{fig:cover_single_mean_3x5_grid}, averaging only five residuals substantially removes the apparent image content and more clearly reveals the shared structure of high-RT methods, providing a more reliable visualization of cover-agnostic watermark signals.

\section{Sensitivity to the Number of Source Images}\label{differentsourcecounts}

As shown in Table~\ref{tab:official-models-rt-comparison}, the overall RT patterns remain consistent across different source-set sizes. High-RT methods remain highly transferable, while HiDDeN and RivaGAN consistently exhibit substantially lower RT, indicating that the observed differences are not specific to a particular source-set size. Considering both evaluation cost and the need to adequately reflect forgery risk, we use 100 source images as the default setting in our main analysis.

\begin{table*}[t]
  \centering
  \caption{
    Sensitivity of residual transferability to the number of source images.
    RT@$s$ is computed using the averaged residual from $s$ source images.
  }
  \label{tab:official-models-rt-comparison}

  \fontsize{9.0}{10.8}\selectfont
  \renewcommand{\arraystretch}{1.02}
  \setlength{\tabcolsep}{3.2pt}

  \begin{tabular}{@{}lcccccc@{}}
    \toprule

    \textbf{Model} &
    \textbf{BitAcc $\uparrow$} &
    \textbf{PSNR (dB) $\uparrow$} &
    \textbf{RT@1 $\downarrow$} &
    \textbf{RT@10 $\downarrow$} &
    \textbf{RT@100 $\downarrow$} &
    \textbf{RT@1000 $\downarrow$} \\

    \midrule

    HiDDeN
      & 0.924$\pm$0.05
      & 33.85$\pm$1.87
      & 0.374$\pm$0.20
      & 0.301$\pm$0.16
      & 0.275$\pm$0.16
      & 0.275$\pm$0.16 \\

    CIN
      & 1.000$\pm$0.00
      & 41.55$\pm$0.90
      & 1.000$\pm$0.00
      & 1.000$\pm$0.00
      & 1.000$\pm$0.00
      & 1.000$\pm$0.00 \\

    MBRS
      & 1.000$\pm$0.00
      & 41.61$\pm$1.69
      & 0.940$\pm$0.08
      & 0.955$\pm$0.07
      & 0.982$\pm$0.04
      & 0.982$\pm$0.04 \\

    RivaGAN
      & 0.919$\pm$0.08
      & 42.05$\pm$0.16
      & 0.081$\pm$0.11
      & 0.088$\pm$0.11
      & 0.082$\pm$0.11
      & 0.085$\pm$0.11 \\

    VINE
      & 1.000$\pm$0.00
      & 35.47$\pm$2.83
      & 0.964$\pm$0.05
      & 0.981$\pm$0.04
      & 0.986$\pm$0.03
      & 0.985$\pm$0.04 \\

    \bottomrule
  \end{tabular}
\end{table*}

\section{Detailed Architectural Analysis}
\label{app:architecture_analysis}
\subsection{Common training-side choices}
\begin{figure}[htbp]
    \centering
    \includegraphics[width=0.24\linewidth]{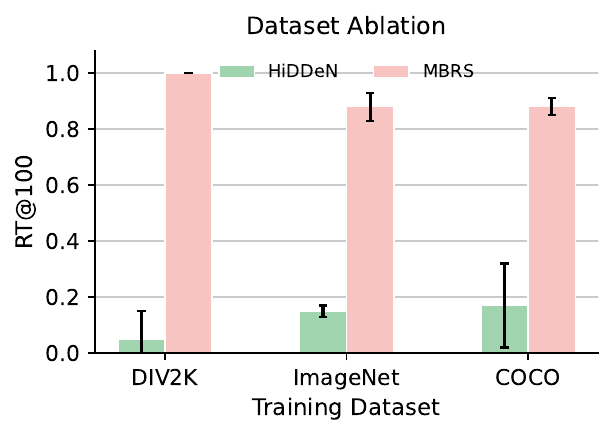}\hfill
    \includegraphics[width=0.24\linewidth]{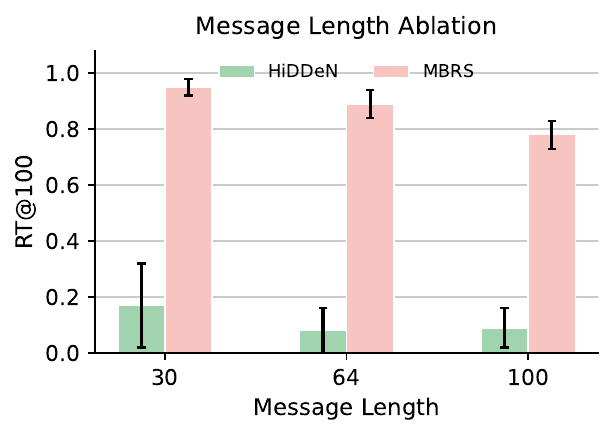}\hfill
    \includegraphics[width=0.24\linewidth]{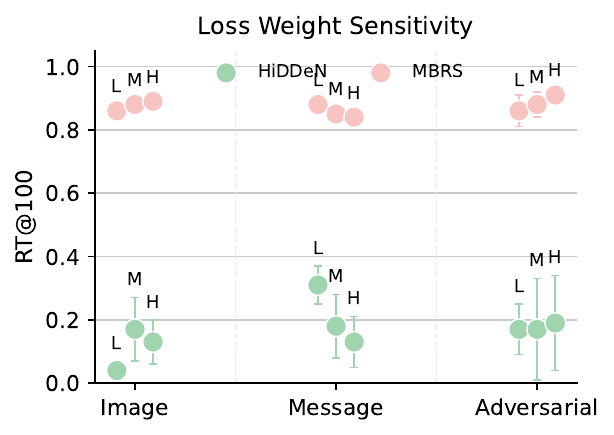}\hfill
    \includegraphics[width=0.24\linewidth]{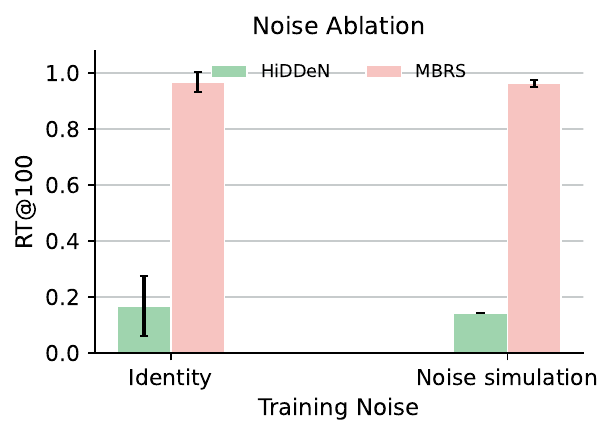}

    \caption{\textbf{Residual transferability (RT@100) under different training settings.}
    We vary the training dataset, message length, loss weight, and robustness-oriented noise simulation while keeping the remaining settings fixed. Results report the mean RT score over three independently trained models with different random seeds, with error bars indicating one standard deviation.}
    \label{fig:training_setting_ablation}
\end{figure}

We test whether common training-side choices can explain the RT gap between HiDDeN~\citep{zhu2018hidden} and MBRS~\citep{jia2021mbrs}. We vary the training dataset, message length, loss weights and robustness-oriented noise simulation while keeping the remaining settings fixed. As shown in Figure~\ref{fig:training_setting_ablation}, these changes shift the exact RT values but do not alter the qualitative behavior: HiDDeN consistently remains in the low-RT regime, whereas MBRS remains highly transferable. This persistence across training settings suggests that the large RT gap is unlikely to be explained by these factors alone, motivating our investigation of architectural differences.

\subsection{Broadcast and GAP as a Coupled Design}
\label{sec:bg_coupling}

To clarify why we treat broadcast and global average pooling (GAP) as a unified Broadcast--GAP
(BG) design rather than two independent architectural choices, we conduct a $2\times2$ ablation on HiDDeN.
Specifically, we independently remove spatial message broadcast from the encoder and GAP from the decoder, while keeping all other components and training settings unchanged. Since the preceding analysis shows that robustness-oriented distortion training does not account for the qualitative RT gap, we omit simulated distortions in the following architectural intervention experiments to reduce optimization difficulty and enable cleaner controlled comparisons. All original and modified models in these experiments are retrained under the same distortion-free setting.

\begin{figure*}[t]
\centering
\captionsetup{hypcap=false}

\begin{minipage}[t]{0.35\textwidth}
    \vspace{0pt}
    \centering

    \includegraphics[width=\linewidth]
    {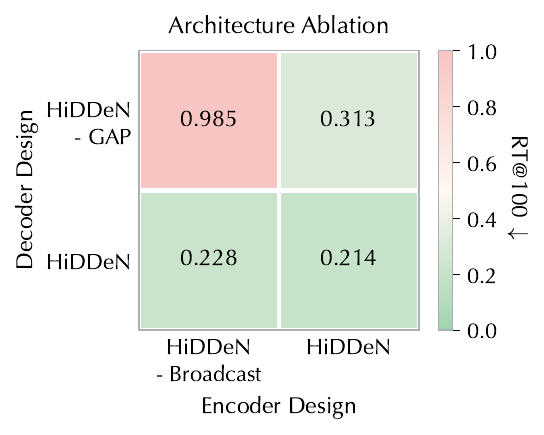}

    \captionof{figure}{
        \textbf{Coupling between broadcast and GAP.}
        Removing both components sharply increases RT.
    }
    \label{fig:bg_architecture_ablation}
\end{minipage}
\hfill
\begin{minipage}[t]{0.60\textwidth}
    \vspace{0pt}
    \centering

    \captionof{table}{
        Identifying architectural sources of low residual transferability.
        RT results are reported as mean $\pm$ standard deviation over three
        random seeds. Lower RT indicates stronger cover dependence.
    }
    \label{tab:cross_family_codesign_rt}

    \fontsize{8.6}{10.2}\selectfont
    \renewcommand{\arraystretch}{1.02}
    \setlength{\tabcolsep}{2.8pt}

    \resizebox{\linewidth}{!}{%
    \begin{tabular}{@{}llccccc@{}}
    \toprule

    \multirow{2}{*}{\textbf{Method}} &
    \multirow{2}{*}{\textbf{Variant}} &
    \multicolumn{2}{c}{\textbf{Watermark Quality}} &
    \multicolumn{3}{c}{\textbf{Residual Transferability}} \\

    \cmidrule(lr){3-4}
    \cmidrule(lr){5-7}

    & &
    \textbf{BitAcc $\uparrow$} &
    \textbf{PSNR $\uparrow$} &
    \textbf{RT@1 $\downarrow$} &
    \textbf{RT@10 $\downarrow$} &
    \textbf{RT@100 $\downarrow$} \\

    \midrule

    \multirow{2}{*}{HiDDeN}
    & Original
    & 0.998
    & 28.23
    & 0.249$\pm$0.180
    & 0.261$\pm$0.135
    & 0.214$\pm$0.135 \\

    & $-$ Broadcast--GAP
    & 1.000
    & 41.47
    & \cellcolor{highrt}\textbf{0.909$\pm$0.084}
    & \cellcolor{highrt}\textbf{0.977$\pm$0.057}
    & \cellcolor{highrt}\textbf{0.985$\pm$0.049} \\

    \addlinespace[2.5pt]

    \multirow{2}{*}{MBRS}
    & Original
    & 1.000
    & 42.32
    & 0.987$\pm$0.029
    & 0.993$\pm$0.023
    & 0.995$\pm$0.020 \\

    & $+$ Broadcast--GAP
    & 0.964
    & 27.07
    & \cellcolor{lowrt}\textbf{0.210$\pm$0.115}
    & \cellcolor{lowrt}\textbf{0.124$\pm$0.109}
    & \cellcolor{lowrt}\textbf{0.098$\pm$0.102} \\

    \addlinespace[2.5pt]

    \multirow{2}{*}{CIN}
    & Original
    & 1.000
    & 41.55
    & 1.000$\pm$0.000
    & 1.000$\pm$0.000
    & 1.000$\pm$0.000 \\

    & $+$ Broadcast--GAP
    & 0.934
    & 21.71
    & \cellcolor{lowrt}\textbf{0.158$\pm$0.157}
    & \cellcolor{lowrt}\textbf{0.103$\pm$0.126}
    & \cellcolor{lowrt}\textbf{0.091$\pm$0.117} \\

    \addlinespace[2.5pt]

    \multirow{2}{*}{VINE}
    & Original
    & 1.000
    & 35.47
    & 0.964$\pm$0.052
    & 0.981$\pm$0.040
    & 0.986$\pm$0.034 \\

    & $+$ Broadcast--GAP
    & 0.600
    & 6.97
    & \multicolumn{3}{c}{\textit{Training failure}} \\

    \addlinespace[2.5pt]

    \multirow{2}{*}{RivaGAN}
    & Original
    & 0.977
    & 42.12
    & 0.071$\pm$0.100
    & 0.073$\pm$0.103
    & 0.077$\pm$0.109 \\

    & Fixed Attention
    & 0.982
    & 42.11
    & \cellcolor{highrt}\textbf{0.922$\pm$0.107}
    & \cellcolor{highrt}\textbf{0.946$\pm$0.094}
    & \cellcolor{highrt}\textbf{0.944$\pm$0.096} \\

    \bottomrule
    \end{tabular}
    }

\end{minipage}

\end{figure*}

As shown in Figure~\ref{fig:bg_architecture_ablation}, the original HiDDeN with both broadcast and GAP achieves an RT@100 of $0.214$.
Removing broadcast alone results in a nearly unchanged RT@100 of $0.228$, while removing GAP alone increases it moderately to $0.313$. In contrast, simultaneously removing both components causes RT@100 to rise sharply to $0.985$.
This pronounced interaction shows that the low-RT behavior cannot be
attributed to either broadcast or GAP independently.
Instead, the two components function as a coupled architectural design: broadcast spatially couples the message with local image features during embedding, while GAP requires the decoder to aggregate spatially distributed evidence for message recovery. We therefore refer to their combination as the \emph{Broadcast--GAP (BG)} design throughout the paper.


\section{Verification Calibration under CoverLock}
\label{sec:false_positive_calibration}

Watermark verification commonly sets its decision threshold using an independent Bernoulli null model. Since CoverLock introduces an additional image-dependent binary code into the verification representation, we examine whether
it alters the null distribution and therefore requires threshold recalibration.

\begin{figure*}[ht]
    \centering

    \begin{minipage}[t]{0.61\textwidth}
        \vspace{0pt}
        \centering

        \includegraphics[width=\linewidth]
        {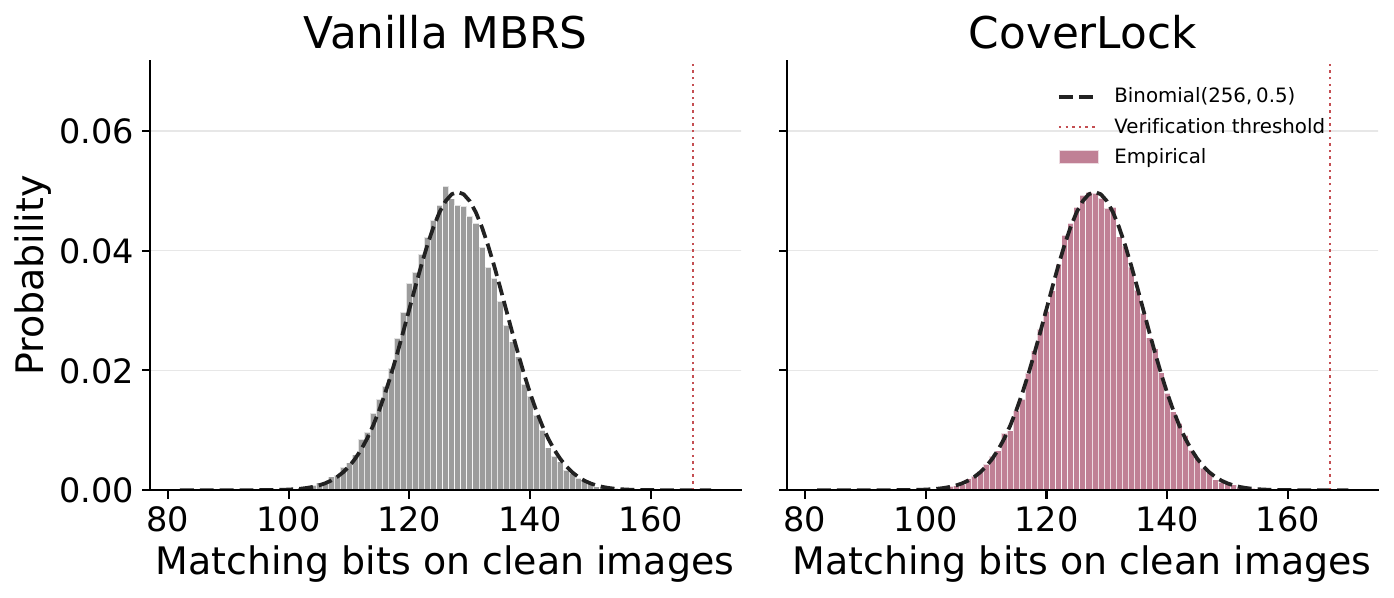}

        \vspace{-3pt}
        \captionof{figure}{
            \textbf{Verification calibration on clean images.}
        }
        \label{fig:false_positive_calibration}
    \end{minipage}
    \hfill
    \begin{minipage}[t]{0.34\textwidth}
        \vspace{0pt}
        \centering

        \captionof{table}{
            Validation ROC-AUC (\%) for classifier backbones.
        }
        \label{tab:appendix_classifier_baselines}

        \vspace{5pt}
        \small
        \setlength{\tabcolsep}{8pt}
        \renewcommand{\arraystretch}{1.15}

        \begin{tabular}{@{}lr@{}}
        \toprule
        \textbf{Backbone} & \textbf{ROC-AUC $\uparrow$} \\
        \midrule
        DINOv2-G + MLP & 72.43 \\
        ResNet-50      & 97.72 \\
        ConvNeXt-Tiny  & \textbf{99.71} \\
        \bottomrule
        \end{tabular}
    \end{minipage}

    \vspace{-5pt}
\end{figure*}

Figure~\ref{fig:false_positive_calibration} compares this theoretical distribution with empirical match counts from
50,000 held-out ImageNet images. Both vanilla MBRS and MBRS with CoverLock-G produce zero false accepts and remain well aligned with the binomial model, with CoverLock-G showing particularly close agreement.

This behavior is consistent with CoverLock's statistical objectives in Section~\ref {app:code_regularization}. In the idealized case where the image-dependent hash $b$ is uniform over $\{0,1\}^L$ and independent of the original decoded bits $\hat{p}$, the decoded message $m'=\hat{p}\oplus b$ is also uniform, yielding a $\mathrm{Binomial}(L,0.5)$ match-count distribution. These objectives encourage the learned binary code to satisfy the statistical conditions underlying the Bernoulli null model, consistent with the close
empirical agreement observed in Figure~\ref{fig:false_positive_calibration}.
Accordingly, the original verification threshold can be retained without additional recalibration after introducing CoverLock.

\section{Classifier Backbone Selection}
\label{app:classifiers}
Inspired by passive detection methods~\citep{wang2026ghostsbeneathtexturestexturerelation,shuai2023locate} for AIGC-generated images, which identify manipulated or synthetic content through discriminative classifiers, we introduce a classifier-based defense as a baseline for detecting residual-transfer forgeries. To select the backbone for the classifier-based defense used in the main experiments, we compare DINOv2-G~\citep{oquab2023dinov2} with an MLP head, ResNet-50~\citep{he2015deep}, and ConvNeXt-Tiny~\cite{liu2022convnet} on MBRS and MSCOCO val2017. We evaluate each classifier on 500 genuine watermarked images and 500 GT@1 residual-transfer forgeries under both clean and distorted settings. An image is accepted only when it passes both the standard watermark verification and the classifier.

As shown in Table~\ref{tab:appendix_classifier_baselines}, ConvNeXt-Tiny provides the strongest overall trade-off between genuine acceptance and forgery rejection among the evaluated backbones, particularly under distortions. We therefore adopt ConvNeXt-Tiny as the classifier backbone in the main experiments.

\end{document}